\documentclass[aps,twocolumn,superscriptaddress,showpacs,dvips]{revtex4}
\usepackage{feynmp}
\usepackage{amssymb}
\usepackage{epsfig}
\usepackage{graphicx}
\usepackage{subfigure}
\usepackage{mathrsfs}
\usepackage{hyperref}
\usepackage{cases}
\usepackage{bbm}

\begin{document}

\title{Order parameter fluctuation and ordering competition in $\mathrm{Ba_{1-x}K_xFe_2As_2}$}

\author{Jing Wang}
\affiliation{Department of Modern Physics, University of Science and
Technology of China, Hefei, Anhui 230026, P.R. China}
\affiliation{Institute for Theoretical Solid State Physics, IFW Dresden,
Helmholtzstr. 20, 01069 Dresden, Germany}
\author{Guo-Zhu Liu}
\affiliation{Department of Modern Physics, University of Science and
Technology of China, Hefei, Anhui 230026, P.R. China}
\author{Dmitry V. Efremov}
\affiliation{Institute for Theoretical Solid State Physics,
IFW Dresden, Helmholtzstr. 20, 01069 Dresden, Germany}
\author{Jeroen van den Brink}
\affiliation{Institute for Theoretical Solid State Physics,
IFW Dresden, Helmholtzstr. 20, 01069 Dresden, Germany}
\affiliation{Institute for Theoretical Physics, TU Dresden,
01069 Dresden, Germany}

\begin{abstract}
The competition among superconductivity, stripe-type magnetic order,
and a new type of $C_4$ symmetric magnetic order in
$\mathrm{Ba_{1-x}K_xFe_2As_2}$ is theoretically studied, focusing on
its impact on the global phase diagram. By carrying out a
renormalization group analysis of an effective field theory, we
obtain the energy-scale dependent flows of all the model parameters,
and then apply the results to understand the observed phase diagram.
On the basis of the renormalization group analysis, we show that the
critical line of nematic order has a negative slope in the
superconducting dome and superconductivity is suppressed near the
magnetic quantum critical point, which are both consistent with
recent experiments. Moreover, we find that, although the observed
$C_4$ symmetric magnetic state could be a charge-spin density wave
or a spin-vortex crystal at high temperatures, charge-spin density
wave is the only stable $C_4$ magnetic state in the low-temperature
regime. Therefore, ordering competition provides a method to
distinguish these two candidate $C_4$ magnetic states.
\end{abstract}


\maketitle

%%%%%%%%%%%%%%%%%%%%%%%%%%%%%Main Body%%%%%%%%%%%%%%%%%%%%%%%%%%%%%%%%%%%%%

\section{Introduction}\label{Sec_intro}

A universal property shared by most known iron-based superconductors
(FeSCs) is the bulk coexistence of two or even more distinct
long-range orders \cite{Hinkov2009NPhys, Paglione2010NPhys,
Stewart2011RMP, Fisher2011RPP, Hirschfeld2011RPP, Basov2011NPhys,
Dai2012NPhys, Chubukov2012ARCMP, Dagotto2013RMP, Fernandes2014NPhys,
Dai2015RMP, Fernandes2017RPP}, such as superconductivity,
stripe-type spin-density-wave (SDW) order, nematic order, and other
possible orders. The competition and coexistence of these orders
leads to a very complicated global phase diagram
\cite{Hinkov2009NPhys, Paglione2010NPhys, Stewart2011RMP,
Fisher2011RPP, Hirschfeld2011RPP, Basov2011NPhys, Dai2012NPhys,
Chubukov2012ARCMP, Dagotto2013RMP, Fernandes2014NPhys, Dai2015RMP,
Fernandes2017RPP}. Acquiring a detailed knowledge of the phase
diagram is an important step towards a better understanding of
FeSCs.

Among the long-range orders competing with superconductivity, a
particular role is played by the nematic order, induced by an
electronic state that spontaneously breaks the $C_4$ (tetragonal)
symmetry of the system down to a $C_2$ (orthogonal) symmetry.
Extensive experiments have confirmed that nematic order exists in
almost all FeSCs \cite{Paglione2010NPhys, Stewart2011RMP,
Fisher2011RPP, Hirschfeld2011RPP, Kim2011PRB, Rotundu2011PRB,
Kasahara2012Nature, Avci2012PRB, Zheng2013NComm}. In most cases, the
nematic order sets in at a temperature $T_\mathrm{n}$ slightly
higher than the critical temperature of magnetic order
$T_\mathrm{m}$ \cite{Basov2011NPhys, Fisher2011RPP,
Fernandes2014NPhys, Fernandes2012PRB, Fernandes2013PRL}. Usually,
the magnetic order is generated by a stripe-type SDW, and possesses
two characteristic vectors $\mathbf{Q}_X = (\pi,0)$ and
$\mathbf{Q}_Y=(0,\pi)$ in the Brillouin zone of the iron square
lattice, which relate to the spin operator $S(\mathbf{r})$ in the
form $S(\mathbf{r}) = M_{X,Y}e^{i\mathbf{Q}_{X,Y} \cdot\mathbf{r}}$
\cite{Fernandes2010PRB,Fernandes2017RPP}. This stripe SDW breaks
the discrete lattice rotational symmetry by selecting out only one
of the two characteristic vectors $\mathbf{Q}_X$ and $\mathbf{Q}_Y$,
preserving the $C_2$ symmetry \cite{Basov2011NPhys,
Chubukov2012ARCMP, Fernandes2014NPhys, Fernandes2017RPP}. Because
the nematic order and SDW order coexist over a large part of the
global phase diagram, it is widely believed \cite{Stewart2011RMP,
Fisher2011RPP, Hirschfeld2011RPP, Basov2011NPhys, Chubukov2012ARCMP,
Dagotto2013RMP, Fernandes2014NPhys, Dai2015RMP, Fernandes2017RPP}
that the nematic order is actually induced by the fluctuation of
magnetic order.

It was unexpected that experiments had found a new type of $C_4$
symmetric magnetic order that preserves the tetragonal symmetry in a
number of hole-doped FeSCs, including
$\mathrm{Ba(Fe_{1-x}Mn_x)_2As_2}$ \cite{Goldman2010PRB},
$\mathrm{Ba_{1-x}Na_xFe_2As_2}$ \cite{Osborn2014NatureComm,
Wang2016PRB}, $\mathrm{Sr_{1-x}K_xFe_2As_2}$
\cite{Osborn2016NaturePhys}, and $\mathrm{Ba_{1-x}K_xFe_2As_2}$
\cite{Hassinger2012PRB, Hardy2015NComm, Allred2015PRB,
Hassinger2016PRB}. This $C_4$ magnetic state is characterized by
biaxial magnetic orders \cite{Lorenzana2008PRL, Eremin2010PRB,
Brydon2010PRB, Giovannetti2011NComm, Fernandes2016PRB}, and the
corresponding spin operator is given by $S(\mathbf{r}) =
M_{X}e^{i\mathbf{Q}_{X}\cdot\mathbf{r}} +
M_{Y}e^{i\mathbf{Q}_{Y}\cdot\mathbf{r}}$ \cite{Fernandes2016PRB,
Schmalian2016PRB}. It has been suggested that this
double-$\mathbf{Q}$ magnetic state has two possible realizations
\cite{Lorenzana2008PRL,Knolle2010PRB, Eremin2010PRB, Brydon2010PRB,
Giovannetti2011NComm, Fernandes2016PRB}: a charge-spin density wave
(CSDW) in which $M_X$ and $M_Y$ are collinear; a spin-vortex crystal
(SVC) in which $M_X$ and $M_Y$ are orthogonal. Since the largest
value of $T_c$ of FeSCs is observed at the proximity of tetragonal
$C_4$ magnetic order \cite{Osborn2014NatureComm, Hardy2015NComm},
there might exist a quantum critical point (QCP) at certain doping
$x_c$ in the superconducting (SC) dome \cite{Fernandes2016PRB}.
After its discovery, the double-$\mathbf{Q}$ structured $C_4$ SDW
state has stimulated a variety of experimental \cite{Goldman2010PRB,
Hassinger2012PRB, Osborn2014NatureComm, Hardy2015NComm,
Allred2015PRB, Osborn2016NaturePhys} and theoretical works
\cite{Lorenzana2008PRL, Eremin2010PRB, Brydon2010PRB,
Giovannetti2011NComm, Kang2011PRB, Anderson2014PRL,
Fernandes2016PRB, Schmalian2016PRB, Anderson2015PRB, Anderson2016PRB}.

\begin{figure}
\centering
\includegraphics[width=3in]{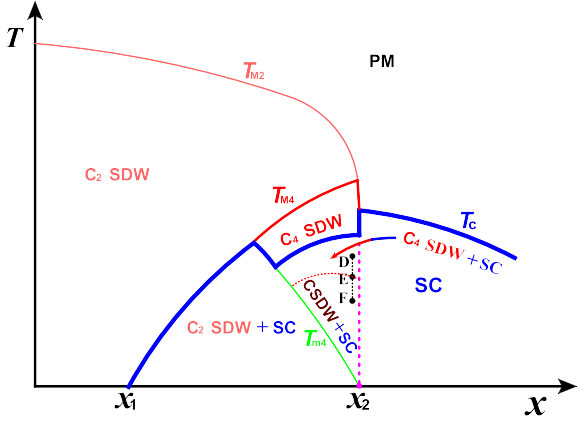}
\vspace{-0.35cm}\caption{(Color online) Schematic global phase
diagram of $\mathrm{Ba_{1-x}K_xFe_2As_2}$ on $(x,T)$ plane
\cite{Hardy2015NComm}, with $x_1$ and $x_2$ being two QCPs. $T_{M2}$
is the transition line between paramagnetic (PM) and $C_2$-symmetric
magnetic phases, and $T_c$ is the SC transition line. $T_{M4}$ is
the transition line between $C_2$- and $C_4$-symmetric magnetic
phases above $T_c$, whereas $T_{m4}$ denotes such a transition line
inside the SC dome. The observed $C_4$ symmetric SDW state could be
either CSDW or SVC. Distinct phases are distinguished by different
values of the model parameters \cite{Fernandes2016PRB,
Schmalian2016PRB} defined in Eq.~(\ref{Eq_L}): (i) $g<0$ and $w=0$,
the system is in PM phase; (ii) $g>\max(0,-w)$ represents the $C_2$
symmetric SDW phase (with nematic order); (iii) $C_4$ symmetric
magnetic state is of SVC-type if $g<0$ and $w>0$, and CSDW-type if
$g<-w$ and $w<0$.} \label{Fig_phase_diagram}
\end{figure}

In this paper, we consider the effects caused by the competition of
superconductivity with both stripe-type $C_2$ symmetric and $C_4$
symmetric magnetic orders in a hole-doped FeSC
$\mathrm{Ba_{1-x}K_xFe_2As_2}$ \cite{Goldman2010PRB,
Hassinger2012PRB, Osborn2014NatureComm, Hardy2015NComm,
Allred2015PRB, Osborn2016NaturePhys}. Recently, B\"{o}hmer \emph{et
al.} \cite{Hardy2015NComm} have experimentally investigated the
global phase diagram of $\mathrm{Ba_{1-x}K_xFe_2As_2}$, and
identified five distinct thermodynamically stable ordered phases,
which are schematically shown in Fig.~\ref{Fig_phase_diagram}. One
can see that the critical line for the nematic order displays a
rather complicated dependence on doping $x$ and temperature $T$: it
decreases with growing $x$ at high $T$, bends backwards to lower $x$
slightly above $T_c$, and eventually exhibits a negative slope after
penetrating into the SC dome. In the narrow doping region in which
the nematic critical line has a positive slope, $T_c$ is moderately
suppressed. Close to the putative magnetic QCP, represented by $x_2$
in Fig.~\ref{Fig_phase_diagram}, there appears on the phase diagram
a region that manifests $C_4$ symmetric SDW state, which occupies
part of the usual $C_2$ symmetric SDW phase and coexists with
superconductivity below $T_c$. In principle, the experimentally
observed $C_4$ SDW state might be a CSDW or SVC type state, which
needs to be clarified theoretically.

Instead of trying to explain the entire phase diagram observed in
Ref.~\cite{Hardy2015NComm}, we perform a more moderate task in this
work. In particular, we will concentrate on the narrow doping region
surrounding the magnetic QCP $x_2$ inside the SC dome and endeavor
to answer the following questions. How to determine whether the
observed $C_4$ magnetic state is of CSDW or SVC type? What is the
scenario that leads to suppresses superconductivity near the
magnetic QCP? Why does the nematic critical line display a negative,
rather than positive, slope in the SC dome? We will address these
issues by investigating the impact of order competition in the
low-$T$ regime.

We study an effective field theory that can be used to describe the
low-energy physics of $\mathrm{Ba_{1-x}K_xFe_2As_2}$ and other
analogous FeSCs \cite{Fernandes2010PRB, Knolle2010PRB,
Eremin2010PRB, Fernandes2012PRB, Fernandes2013PRL,
Fernandes2017RPP}. When the parameters used in this model take
different values, the system might be in the paramagnetic state,
$C_4$ tetragonal SDW state, or $C_2$ symmetric magnetic state. The
fluctuations of the associated order parameters and the competition
between distinct orders can qualitatively alter the magnitudes and
even the sign of the model parameters, which would drive phase
transitions and reshape the global phase diagram. After analyzing
the competition between nematic and SC orders by means of
renormalization group (RG) method, we find that the slope of nematic
critical line is always negative in the SC dome. We also extract the
$T$-dependence of superfluid density from RG results, which clearly
shows that superconductivity is suppressed near magnetic QCP.
Moreover, we infer from the RG results that, although CSDW and SVC
state are both possible in the high-$T$ regime, the CSDW state is
the only stable one in the low-$T$ regime, which provides a
promising way to specify the true nature of the observed $C_4$
symmetric SDW state.

The rest of paper is organized as follows. In
Sec.~\ref{Sec_eff_action}, we present the effective field theory for
the ordering competition and then derive the coupled flow equations
of all the model parameters by performing perturbative RG
calculations. In Sec.~\ref{Sec_discussions}, we numerically solve
the equations and apply the RG solutions to understand several
important features of the global phase diagram of
$\mathrm{Ba_{1-x}K_xFe_2As_2}$ observed in recent experiments. The
Sec.~\ref{Sec_discussions_2} is followed to present some
discussions. In Sec.~\ref{Sec_summary}, we present a brief summary
of the results.

\section{Effective theory and RG analysis}\label{Sec_eff_action}

Many of the basic properties of $\mathrm{Ba_{1-x}K_xFe_2As_2}$ can
be described by a three-band model that contains one hole pocket at
the center of Brillouin zone $\mathbf{Q}_\Gamma = (0,0)$ and two
electron pockets centered at two specific momenta
$\mathbf{Q}_X=(\pi,0)$ and $\mathbf{Q}_Y = (0,\pi)$
\cite{Knolle2010PRB, Eremin2010PRB, Fernandes2010PRB,
Fernandes2012PRB, Fernandes2013PRL, Fernandes2016PRB,
Schmalian2016PRB, Fernandes2017RPP}. The microscopic model is
written as \cite{Fernandes2013PRL, Fernandes2017RPP}
\begin{eqnarray}
H=\sum_{\mathbf{k},i\in(X,Y,\Gamma)}\varepsilon_{\mathbf{k},i}
c^\dagger_{\mathbf{k}\sigma,i}c_{\mathbf{k}\sigma,i} + H_4,
\end{eqnarray}
with the interacting term $H_4$ is given by
\begin{eqnarray}
H_4 &=& \sum_{\mathbf{k},i\in(X,Y)}\frac{U_3}{2}
\left(c^\dagger_{\mathbf{k}\alpha,\Gamma}
c^\dagger_{\mathbf{k}\gamma,\Gamma}
c_{\mathbf{k}\delta,i}c_{\mathbf{k}\beta,i} + \mathrm{h.c.}\right)
\delta_{\alpha\beta}\delta_{\gamma\delta}\nonumber\\
&& + \sum_{\mathbf{k},i\in(X,Y)}U_1
c^\dagger_{\mathbf{k}\alpha,\Gamma} c^\dagger_{\mathbf{k}\gamma,i}
c_{\mathbf{k}\delta,i}c_{\mathbf{k}\beta,\Gamma}
\delta_{\alpha\beta}\delta_{\gamma\delta}.
\end{eqnarray}
Here, $U_1$ and $U_3$ represent density-density interaction and the
pair hoping interaction, respectively. They are responsible for the
formation of superconductivity and SDW state \cite{Fernandes2013PRL,
Chubukov2008PRB, Maiti2010PRB}. The magnetic structure can be
described by two order parameters $\mathbf{M}_X$ and $\mathbf{M}_Y$,
corresponding to the ordering vectors $\mathbf{Q}_X=(\pi,0)$ and
$\mathbf{Q}_Y=(0,\pi)$, which are defined as $\mathbf{M}_j =
\sum_{\mathbf{k}}c^\dagger_{\Gamma,\mathbf{k}
\alpha}\vec{\sigma}_{\alpha\beta} c_{j,\mathbf{k} +
\mathbf{Q}_j\beta}$ with $j = X,Y$ \cite{Fang2008PRB, Xu2008PRB,
Fernandes_Schmalian2010PRL, Fernandes2012PRB, Fernandes2013PRL}.
Both the $C_2$ and $C_4$ symmetric magnetic orders are modeled by
the following Ginzburg-Landau free energy \cite{Fernandes2016PRB,
Schmalian2016PRB}
\begin{eqnarray}
f[\mathbf{M_X},\mathbf{M_Y}] &=& a_m(\mathbf{M}^2_X +
\mathbf{M}^2_Y) +
\frac{u}{2}(\mathbf{M}^2_X+\mathbf{M}^2_Y)^2\nonumber \\
&&-\frac{g}{2}(\mathbf{M}^2_X-\mathbf{M}^2_Y)^2 +
2w(\mathbf{M}_X\cdot\mathbf{M}_Y)^2.\nonumber
\end{eqnarray}
As illustrated in Ref.~\cite{Fernandes2016PRB}, the term
$2w(\mathbf{M}_X \cdot \mathbf{M}_Y)^2$ can be rewritten by using an
identity:
\begin{eqnarray}
(\mathbf{M}_X \cdot \mathbf{M}_Y)^2 &=& \frac{1}{4}(\mathbf{M}^2_X +
\mathbf{M}^2_Y)^2 - \frac{1}{4}(\mathbf{M}^2_X - \mathbf{M}^2_Y)^2
\nonumber \\
&&-(\mathbf{M}_X \times \mathbf{M}_Y)^2.
\end{eqnarray}
Upon carrying out a Hubbard-Stratonovich transformation followed by
an integration over all the fermionic degrees of freedom, one can
obtain an effective field theory \cite{Fernandes2013PRL,
Fernandes2016PRB} for the interplay of SDW magnetic and SC orders in
the vicinity of magnetic QCP:
\begin{eqnarray}
\mathcal{L} &=& \frac{1}{2}(\partial_\mu M_X)^2 +
\frac{1}{2}(\partial_\mu
M_Y)^2 + a_m\left(M_X^{2} + M_Y^{2}\right)\nonumber \\
&&+\frac{(u + w)}{2}\left(M_X^{2} + M_Y^{2}\right)^{2} -
\frac{\left(g + w\right)}{2}
\left(M_X^{2} - M_Y^{2}\right)^{2}\nonumber\\
&&+\partial_\mu \Delta^\dagger\partial_\mu\Delta + a_s\Delta^2(k) +
\frac{u_s}{2}\Delta^4(k) +
\frac{\varphi^{2}}{2w}+\mathcal{L}_A\nonumber \\
&& -2\varphi M_X M_Y + \lambda(M^2_X + M^2_Y)\Delta^2
+\lambda_{\Delta A}\Delta^2 A^2,\label{Eq_L}
\end{eqnarray}
where $\Delta$ is the SC order parameter. Here we use a positive
parameter $\lambda$ to characterize the repulsive interaction
(competition) between SC and magnetic orders. In order to evaluate
the superfluid density, we have introduced a gauge potential
$\mathbf{A}$ via the standard minimal coupling
\cite{Halperin1974PRL} with $\mathcal{L}_A =
-\frac{1}{4}(\partial_\mu A_\nu -\partial_\nu A_\mu)^2$. This model
contains eight fundamental parameters $a_m$, $a_s$, $u_s$, $w$, $u$,
$g$, $\lambda$, and $\lambda_{\Delta A}$, which are constants at the
mean-field level, but all become cutoff dependent due to
interactions.

The transition lines for SDW and SC orders are determined by taking
$a_m = 0$ and $a_s = 0$ respectively. For $s^{+-}$-wave
superconductors, we employ the relationship $\Delta_\Gamma = -
\sqrt{2}\Delta_{X,Y} = \Delta$ \cite{Fernandes2013PRL,
Chubukov2012ARCMP, Fernandes2017RPP}. An Ising-type nematic order
is induced by the magnetic order, and represented by a term of the
form $M^2_X - M^2_Y$ \cite{Fernandes2012PRB, Chubukov2012ARCMP,
Fernandes2017RPP}. The property of $C_4$ magnetic order is
determined by the parameter $w$ \cite{Fernandes2016PRB,
Schmalian2016PRB}. In the SC dome, the SC order parameter develops a
nonzero mean value, i.e., $\langle\Delta\rangle = V_0 =
\sqrt{-a_s/u_s}$ near the magnetic QCP $x_2$.

The effective model (\ref{Eq_L}) displays different states when the
model parameters take various values \cite{Fernandes2016PRB,
Schmalian2016PRB}. (i) If $g<0$ and $w=0$, the effective model is in
paramagnetic (PM) phase; (ii) The case of $g>\max(0,-w)$ corresponds
to the $C_2$ SDW phase (with nematic order); (iii) The $C_4$
symmetric magnetic state is of SVC-type if $g<0$ and $w>0$, and
CSDW-type if $g<-w$ and $w<0$. Once some of these parameters are
altered by external forces, such as doping, magnetic field, and
pressure, the system would undergo transitions between distinct
phases. However, the quantum fluctuations of order parameters and
the interaction between different order parameters can also lead to
remarkable changes of model parameters, and as such drive phase
transitions. In the next section, we will study the RG flows of
these parameters and examine how they are influenced by order
parameter fluctuation and ordering competition. The main results are
schematically illuminated in Fig.~\ref{Fig_phase_diagram} and the
detailed derivations and discussions are given in the following.

To proceed, we perform a RG analysis of the effective theory
(\ref{Eq_L}). Our focus is on the behavior of the system at low $T$
and in the vicinity of magnetic QCP. Within this region, the quantum
fluctuations of SC order parameter can result in drastic effects
even in the SC phase. For the complex SC order parameter
$\Delta(\mathbf{r})$, there are two sorts of fluctuations
\cite{Varma2002JLTP, Zwerger2004PRL, Podolsky2011PRB,
Podolsky2012PRB, Pollet2012PRL, Bloch2012Nature, Varma2013PRB,
Pekker2015ARCMP}, namely the phase fluctuation and amplitude
fluctuation. The former fluctuation is gapless and corresponds to
the Nambu-Goldstone mode induced by continuous gauge symmetry
breaking. This mode does not play any role in the SC state because
it is absorbed by the vector gauge boson via the Anderson-Higgs
mechanism. The latter one, known as Higgs mode in a locally gauge
invariant superconductor, is found by both theoretical and
experimental works to result in observable effects
\cite{Varma2002JLTP, Zwerger2004PRL, Podolsky2011PRB,
Podolsky2012PRB, Pollet2012PRL, Bloch2012Nature, Varma2013PRB,
Pekker2015ARCMP}, and hence should be seriously considered
\cite{Kleinert2003NPB, Wang2014PRD}. In order to capture the quantum
fluctuation of SC order parameter around its mean value $\langle
\Delta \rangle$, we define two new fields $h$ and $\eta$ by
\cite{Kleinert2003NPB, Wang2014PRD}
\begin{eqnarray}
\Delta = V_0+\frac{1}{\sqrt{2}}(h+i\eta),\label{Eq_Delta_eta_h}
\end{eqnarray}
where $\langle h\rangle = \langle \eta \rangle = 0$. The fields $h$
and $\eta$ stand for the Higgs mode and Nambu-Goldstone mode,
respectively. We substitute Eq.~(\ref{Eq_Delta_eta_h}) into the
effective Lagrangian density $\mathcal{L}$ (\ref{Eq_L}), and obtain
the following new effective Lagrangian density:
\begin{widetext}
\begin{eqnarray}
\mathcal{L}_{\mathrm{eff}} &=& \frac{1}{2}(\partial_\mu
M_X)^2+\alpha_{X}M^2_X + \frac{\beta_{X}}{2}M_X^{4} +
\frac{1}{2}(\partial_\mu M_Y)^2+\alpha_{Y}M^2_Y +
\frac{\beta_{Y}}{2}M_Y^{4} + \frac{1}{2}(\partial_\mu h)^2 +
\alpha_h h^2 + \frac{\beta_h}{2}h^4 +
\gamma_h h^3 \nonumber \\
&&+\left(\mathcal{L}_{A}+\frac{\alpha_A}{2}A^2\right) +
\alpha_\varphi \varphi^{2} + \gamma_{\varphi X Y}\varphi M_XM_Y +
\gamma_{X^2h}M^2_X h + \gamma_{Y^2 h}M^2_Y h +
\gamma_{hA^2}h A^2 + \lambda_{X h}M^2_Xh^2\nonumber\\
&& +\lambda_{Y h}M^2_Y h^2 + \lambda_{XY}M_X^{2} M_Y^{2} +
\lambda_{hA}h^2A^2.\label{Eq_effective_L}
\end{eqnarray}
The gapless Nambu-Goldstone model $\eta$ naturally disappears after
invoking the Anderson-Higgs mechanism. However, the Higgs mode $h$
remains in the above effective model, and couple directly to the
magnetic order parameters $M_{X,Y}$ and also to vector potential
$A$. The originally massless gauge field $A$ acquires a finite mass
$\alpha_A$ after absorbing $\eta$. Moreover, in the above Lagrangian
density we have introduced a number of new parameters that are
related to the model parameters defined in (\ref{Eq_L}) by the
following relations:
\begin{eqnarray}
\left.\begin{array}{ll} \alpha_{X}=\alpha_{Y}\equiv a_m -
\frac{\lambda a_s}{u_s},\,\,\,\,\,\alpha_A\equiv \frac{-2\lambda_{\Delta
A}a_s}{u_s}, \,\,\,\,\,\alpha_h\equiv - a_s,\,\,\,\,\,\alpha_\varphi \equiv
\frac{1}{2w},\,\,\,\,\,\beta_{X}=\beta_{Y}\equiv
u - g,\,\,\,\,\,\beta_h\equiv\frac{u_s}{4},\\
\gamma_h\equiv\frac{\sqrt{-2a_su_s}}{2},\,\,\,\,\,\,\,\gamma_{hA^2} \equiv
\sqrt{\frac{-2\lambda^2_{\Delta A}a_s}{u_s}},\,\,\,\,\,\,\,\gamma_{\varphi X
Y}\equiv-2,\,\,\,\,\,\,\,\gamma_{X^2h} = \gamma_{Y^2h}\equiv
\sqrt{\frac{-2\lambda^2a_s}{u_s}},\\
\lambda_{XY}\equiv u + g + 2w,\,\,\,\,\,\,\,
\lambda_{hA}\equiv\frac{\lambda_{\Delta A}}{2},\,\,\,\,\,\,\,\lambda_{X h} =
\lambda_{Y h}\equiv\frac{\lambda}{2}.\label{Eq_para_trans}
\end{array}\right.
\end{eqnarray}
Using these relations, we can derive the flow equations of
fundamental parameters by calculating the effective parameters
$\alpha_X$, $\alpha_h$, $\alpha_\varphi$, $\beta_X$, $\beta_h$,
$\lambda_{XY}$, $\lambda_{Xh}$, and $\lambda_{hA}$. By performing
perturbative expansion in powers of small coupling parameters
\cite{Shankar1994RMP} and utilizing $\dot{u}$ to denote the
derivative of $u$ with respect to the varying length scale $l$, we
arrive at the following flow equations with the help of the
identifies given by Eq.~(\ref{Eq_para_trans}) \cite{Wang2014PRD}:
\begin{eqnarray}
\left.\begin{array}{ll} \dot{a}_m = 2\left(a_m-\frac{\lambda
a_s}{u_s}\right)+\frac{1}{4\pi^2}\left\{\frac{\lambda}{2}(1+2a_s) +
\frac{8\lambda^2a^2_s}{u_s}+\left[1-2\left(a_m-\frac{\lambda
a_s}{u_s}\right)\right]
\left[2(2u-g)+\frac{4\lambda^2a_s}{u_s}\right]\right\} \vspace{0.2cm} \\
\hspace{0.77cm} + \left(\frac{\lambda}{u_s}\dot{a}_s +
\frac{a_s}{u_s}\dot{\lambda}-\frac{a_s\lambda}{u_s^2}\dot{u}_s\right),\vspace{0.35cm}
\\
\dot{a}_s = 2a_s-\frac{1}{12\pi^2}
\left\{\frac{27a_su_s}{2}(1+4a_s) +
\frac{12a_s\lambda^2}{u_s}\left[1-4\left(a_m -
\frac{a_s\lambda}{u_s}\right)\right] +
3\lambda\left[1-2\left(a_m-\frac{a_s\lambda}
{u_s}\right)\right]\right. \vspace{0.2cm}\\
\hspace{0.77cm}\left.+\frac{9u_s}{4}(1+2a_s) + 3\lambda_{\Delta
A}\left(1+\frac{2a_s\lambda_{\Delta A}}
{u_s}\right)+\frac{32a_s\lambda^2_{\Delta
A}}{u_s}\left(1+\frac{4a_s\lambda_{\Delta A}} {u_s}\right)\right\},\vspace{0.35cm}
\\
\dot{u}_s = u_s+\frac{1}{\pi^2}\left\{-\frac{9u^2_s}{4}(4a_s +
1)+2\lambda^2 \left[4\left(a_m -
\frac{a_s\lambda}{u_s}\right)-1\right] +
54a_su^2_s(1+6a_s)-\frac{4\lambda^2_{\Delta A}}{3}
\left(\frac{4a_s\lambda_{\Delta A}}{u_s}+1\right)\right. \vspace{0.2cm} \\
\hspace{0.77cm}\left.+\frac{32a_s\lambda^3}{u_s}\left(1 +
\frac{6a_s\lambda}{u_s}\right)+\frac{11072a_s \lambda^3_{\Delta
A}}{35u_s}\left(1+\frac{6a_s\lambda_{\Delta A}}
{u_s}\right)\right\},\vspace{0.35cm} \\
\dot{u} = u+w+\frac{1}{2\pi^2}\left\{\left[9(u-g)^2 +
3(u+g+2w)(u-g)+5(u+g+2w)^2+12w(u-g)\right]\left[4\left(a_m-
\frac{a_s\lambda}{u_s}\right)-1\right]\right. \vspace{0.2cm} \\
\hspace{0.77cm}-\frac{3\lambda^2}{8}(4a_s+1) +
\frac{24a_s\lambda^2(u-g)}{u_s}\left[1-2\left(2\left(a_m-
\frac{a_s\lambda}{u_s}\right) - a_s\right)\right] +
4w(u+g+2w)\left[4\left(a_m- \frac{a_s\lambda}{u_s}\right) - 1\right] \vspace{0.2cm} \\
\hspace{0.77cm}\left.+\frac{8a_s\lambda^2(u+g+2w)}{u_s}
\left[1-4\left(a_m-\frac{a_s\lambda}
{u_s}\right)+2a_s\right]+\frac{4a_s\lambda^3}{u_s}\left[1 -
2\left(a_m-\frac{a_s\lambda} {u_s}\right) +
4a_s\right]\right\}-\dot{w}, \vspace{0.35cm} \\
\dot{g} =g+w-\frac{1}{2\pi^2}\left\{\left[9(u-g)^2 -
3(u+g+2w)(u-g)-3(u+g+2w)^2-12w(u-g)\right]\left[4\left(a_m-
\frac{a_s\lambda}{u_s}\right)-1\right]\right. \vspace{0.2cm}\\
\hspace{0.77cm}-\frac{\lambda^2}{8}(4a_s+1) +
\frac{24a_s\lambda^2(u-g)}{u_s}\left[1-2\left(2\left(a_m-
\frac{a_s\lambda}{u_s}\right)-a_s\right)\right] +
4w(u+g+2w)\left[4\left(a_m-\frac{a_s\lambda}{u_s}\right)-1\right] \vspace{0.2cm}\\
\hspace{0.77cm}\left.-\frac{8a_s\lambda^2(u+g+2w)}{u_s}
\left[1-4\left(a_m - \frac{a_s\lambda}{u_s}\right) + 2a_s\right]
+\frac{4a_s\lambda^3}{u_s}\left[1-2\left(a_m - \frac{a_s\lambda}
{u_s}\right) + 4a_s\right]\right\}-\dot{w},\vspace{0.35cm}
\\
\dot{\lambda} = \lambda+\frac{1}{\pi^2}\left\{\frac{4a_s\lambda^3}{u_s}\left[1 -
4\left(a_m-\frac{a_s\lambda}{u_s}\right) +
2a_s\right]+\lambda(2u-g+3w)\left[4\left(a_m -
\frac{a_s\lambda}{u_s}\right)-1\right]\right. \vspace{0.2cm}\\
\hspace{0.77cm}+\frac{16a_s\lambda^2(2u-g+w)}{u_s}
\left[1-6\left(a_m-\frac{a_s\lambda}{u_s}\right)\right] +
6a_s\lambda^2\left[1-2\left(a_m-\frac{a_s\lambda}
{u_s}\right)+4\alpha_s\right] \vspace{0.2cm}\\
\hspace{0.77cm}\left.-\frac{3u_s\lambda}{8}(4a_s+1) +
2\lambda^2\left[2\left(a_m-\frac{a_s\lambda}{u_s}\right) - 2a_s -
1\right]+9a_su_s\lambda(1+6a_s)\right\},\vspace{0.35cm}
\\
\dot{\lambda}_{\Delta A} = \lambda_{\Delta
A}+\frac{1}{3\pi^2}\left\{27a_su_s\lambda_{\Delta
A}(1+6a_s)+\frac{4\lambda^2_{\Delta A} (16a_s\lambda_{\Delta
A}-3u_s)}{u_s}\left[1+2a_s\left(1+\frac{2\lambda_{\Delta A}}
{u_s}\right)\right]\right. \vspace{0.2cm}\\
\hspace{0.77cm}\left.-\frac{9u_s\lambda_{\Delta A}}{8}(4a_s + 1) +
36a_s\lambda^2_{\Delta A} \left[1+2a_s\left(2+\frac{\lambda_{\Delta
A}}{u_s}\right)\right]\right\},\vspace{0.35cm}
\\
\dot{w}= \frac{w^2}{\pi^2}\left[1-4\left(a_m-\frac{\lambda
a_s}{u_s}\right)\right].
\end{array}\right.\label{Eq_Renormalized_RG_QCP}
\end{eqnarray}
\end{widetext}

\section{Comparison with experiments}\label{Sec_discussions}

In this section, we will compare the RG results with recent
experiments. We first numerically solve the self-consistently
coupled RG equations, and then manage to understand a number of
important features observed by B\"{o}hmer \emph{et al.} in
$\mathrm{Ba_{1-x}K_xFe_2As_2}$ \cite{Hardy2015NComm}. We are
particularly interested in the doping dependence of nematic critical
line in the SC dome, the suppression of superconductivity observed
at the magnetic QCP, and the nature of the observed $C_4$ symmetric
magnetic order, which will be studied one by one based on the RG
solutions.

As can be seen from the phase diagram presented in
Fig.~\ref{Fig_phase_diagram}, the magnetic and SC orders are assumed
to coexist over a finite region, with $x_2$ being the magnetic QCP.
Such a coexistence can be realized if the bare values of model
parameters satisfy the constraint \cite{She2010PRB, Vorontsov,
Fernandes2010PRB, Schmalian2010PRB, Fernandes2013PRL, Wang2013NJP}
$\lambda < \sqrt{u_s[(u+w) - (g+w)]} = \sqrt{u_s(u-g)}$. For
simplicity, we will only consider the low-$T$ region in the close
vicinity of the magnetic QCP inside the SC dome. In addition, the
external field $A$ is assumed to be weak, but the basic conclusion
does not depend on this assumption.

\subsection{Slope of nematic critical line in SC dome}\label{Subsec_nematic}

The nematic line is not shown apparently in
Fig.~\ref{Fig_phase_diagram}. However, the $C_2$ magnetic phase is
always accompanied (even preempted) by a nematic phase with a
critical temperature $T_n$ higher than that of $C_2$ magnetic order
\cite{Basov2011NPhys, Chubukov2012ARCMP, Fernandes2013PRL,
Fernandes2014NPhys}. Hence $T_{m4}$ is also a nematic critical line.

A known fact is that a long-range order can always be destroyed by
thermal fluctuation at sufficiently high $T$. In a system containing
two or more distinct orders, the competition between these orders
might destroy some specific order at very low $T$. As a result, the
critical line on the $(x,T)$ phase diagram for this specific order
has a positive slope in the low-$T$ region, which is often called
back-bending behavior. Interestingly, such back-bending behavior has
been observed in some high-$T_c$ cuprate superconductor
\cite{Vishik, Hashimotonp2014, Hashimotonm2015} and FeSCs
\cite{Nandi2010PRL}. In cuprate Bi$_2$Sr$_2$CaCu$_2$O$_{8+\delta}$,
a pseudogap exists above $T_c$ on the $(x,T)$ phase diagram. This
pseudogap decreases rapidly with growing doping $x$, so its critical
line exhibits a negative slope above $T_c$. However, after entering
into the SC dome, the critical line for pseudogap was found to bend
backwards to lower doping, and thus displays a positive slope in the
low-$T$ region \cite{Vishik, Hashimotonp2014, Hashimotonm2015}. A
simimar behavior was also observed in
$\mathrm{Ba(Fe_{1-x}Co_x)_2As_2}$ by Nandi \emph{et al.}
\cite{Nandi2010PRL}. In this case, it is the nematic order that is in
strong competition with SC order. The nematic critical line has a
negative slope on $(x,T)$ phase diagram above $T_c$, but displays a
positive slope below $T_c$ \cite{Nandi2010PRL}. A common feature
observed in these two compounds is that the critical line for the
order competing with superconductivity has a positive slope in the
low-$T$ region. While a convincing theoretic explanation for the
back-bending of pseudogap critical line in
Bi$_2$Sr$_2$CaCu$_2$O$_{8+\delta}$ is lacking, a recent RG work
reproduced the back-bending of nematic critical line by studying the
competition between nematic and SC orders in
$\mathrm{Ba(Fe_{1-x}Co_x)_2As_2}$ \cite{Wang2015PRB}. Different from
$\mathrm{Ba(Fe_{1-x}Co_x)_2As_2}$, the nematic critical line has a
negative slope in the SC dome of $\mathrm{Ba_{1-x}K_xFe_2As_2}$
\cite{Hardy2015NComm} despite the presence of ordering competition.

\begin{figure}
\includegraphics[width=3.65in]{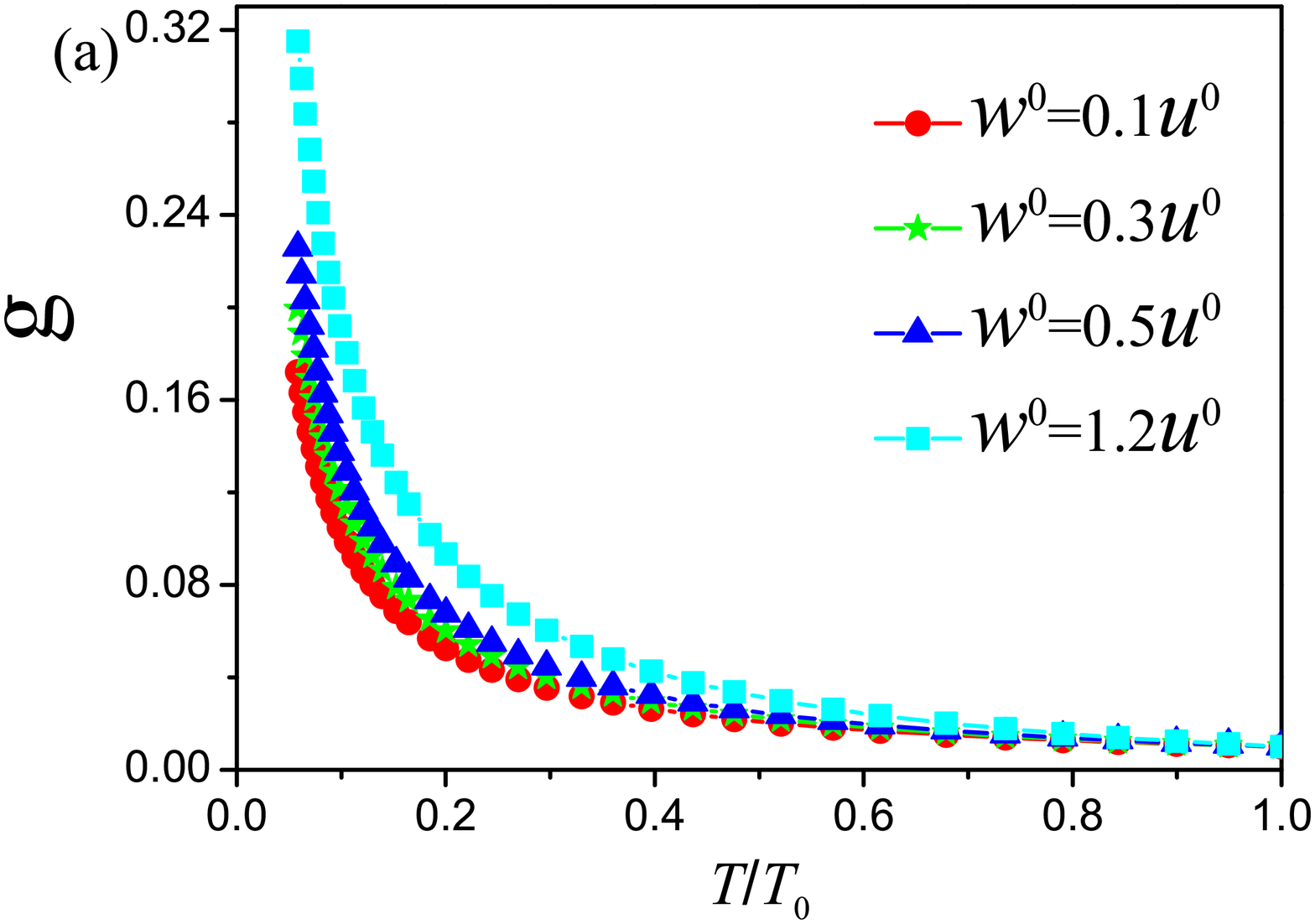}\vspace{-0.6cm}
\includegraphics[width=3.65in]{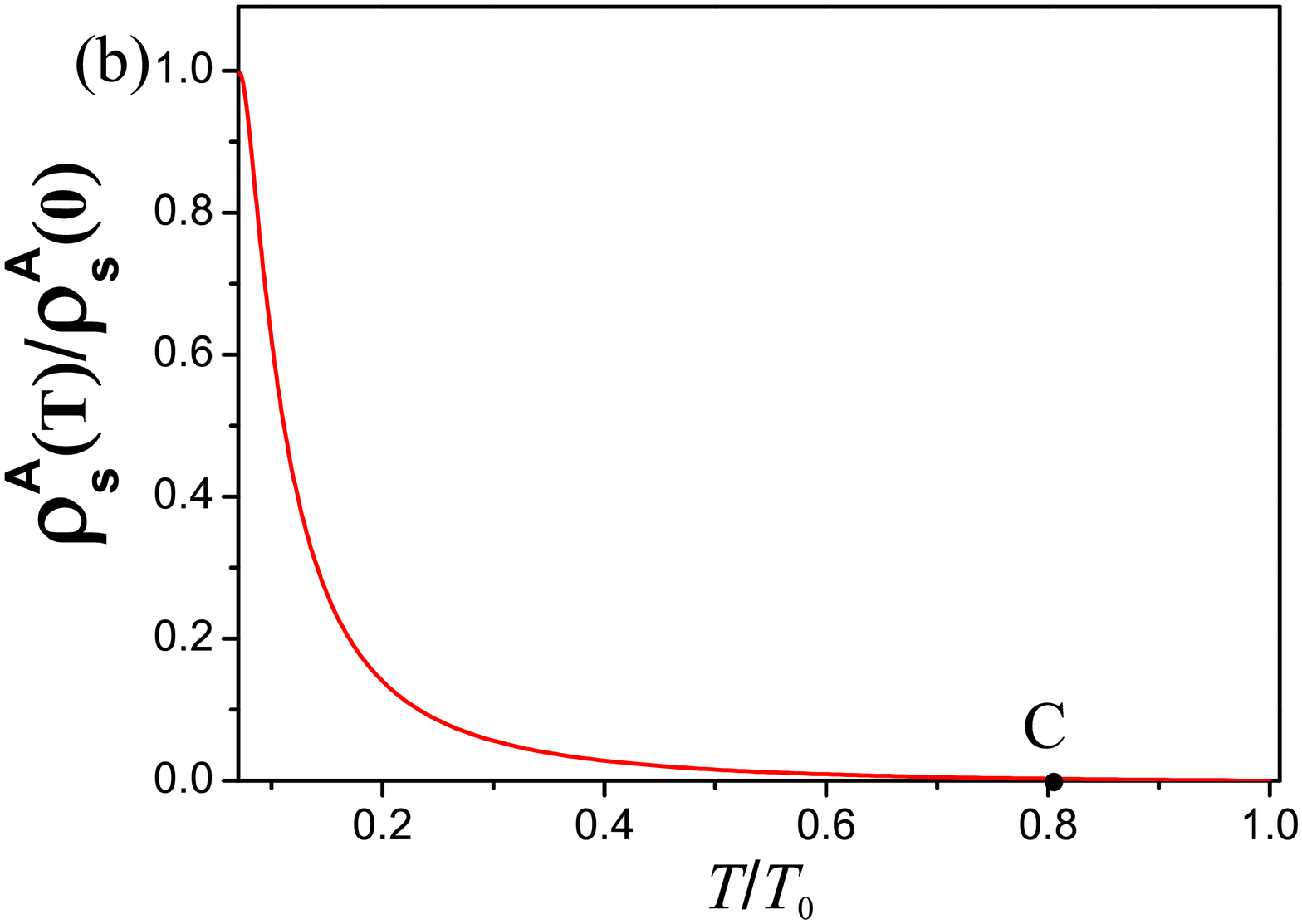}
\vspace{-0.82cm}\caption{(Color online) (a) $T$-dependence of
parameter $g$ at several different values of $w_0$. Calculations
confirm that the positive parameter $g$ never becomes negative as
$T$ decreases. (b) $T$-dependence of $\rho^A_s(T)/\rho^A_s(0)$. The
quantity $\rho^A_s(T=0)$ is temperature independent at sufficiently
low temperatures in the absence of ordering competition. It is
apparent that $\rho^A_s(T)$ vanishes at certain point $C$, where
$T\approx 0.8T_0$.} \label{Fig_Tem_dependece}
\end{figure}

In $\mathrm{Ba(Fe_{1-x}Co_x)_2As_2}$, there is only $C_2$ symmetric
magnetic order, induced by a stripe-type SDW state. As discussed in
Ref.~\cite{Fernandes2013PRL}, the existence of nematic order is
tuned by the quadratic term $-g(M^2_X - M^2_Y)^2$. The system can
stay either in the PM phase, or in one of the $M_X$ and $M_Y$
magnetically ordered phases. The former case corresponds to a state
in which $g < 0$ and no nematic order exists. In the latter case, $g
> 0$ and hence the system exhibits a nematic order. Both of these
two possibilities can be realized at low $T$. When the competition
between nematic and SC orders is sufficiently strong, it is in
principle possible for the nematic order to be suppressed in the
low-$T$ region, leading to a positive slope of nematic critical line
in the SC dome \cite{Wang2015PRB}.

We now use the RG solutions to judge whether the nematic critical
line has a positive or negative slope in the SC dome of
$\mathrm{Ba_{1-x}K_xFe_2As_2}$. In the effective field model given
by Eq.~(\ref{Eq_L}), the relation between $g$ and $w$ determines
whether the nematic order is present or not. As pointed out
previously in Refs.~\cite{Fernandes2014NPhys, Fernandes2012PRB,
Wang2015PRB}, when $g > \max(0,-w)$, only one of the two order
parameters $M_X$ and $M_Y$ develops a finite mean value due to
tetragonal symmetry breaking, which is a clear signature for the
existence of a nematic order. On the other hand, we have $\langle
M_X \rangle = \langle M_Y \rangle$ if $g<\max(0,-w)$, which implies
the absence of nematic order \cite{Fernandes2012PRB,
Fernandes2014NPhys, Wang2015PRB}. This property will be used to
judge whether the nematic critical line bends back.

\begin{figure}
\includegraphics[width=1.7in]{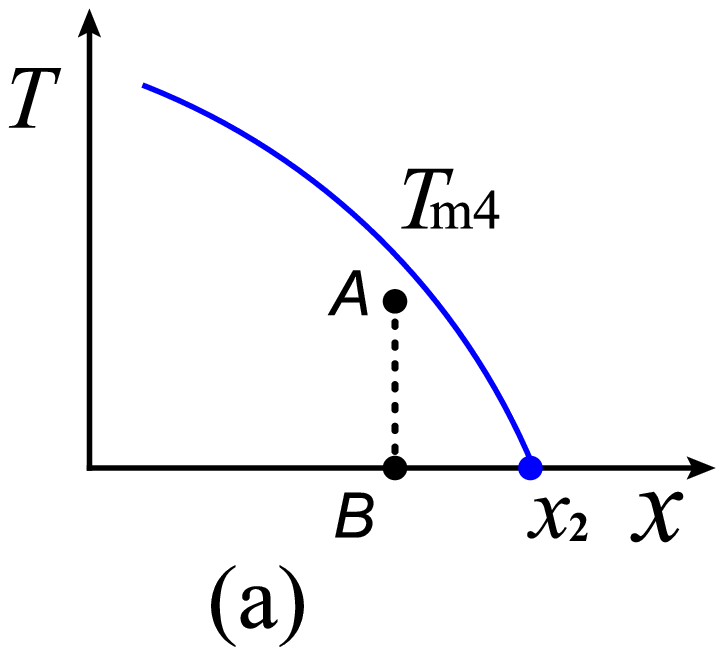}\hspace{-0.3cm}
\includegraphics[width=1.7in]{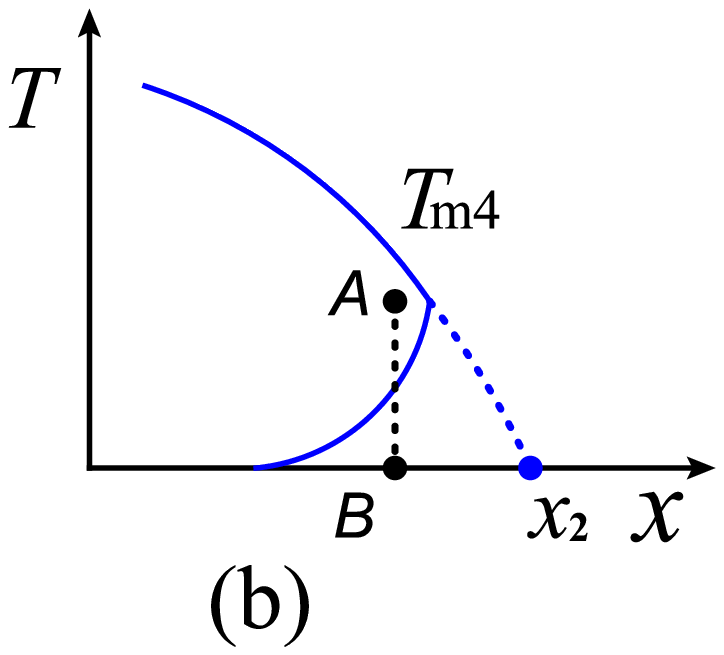}
\vspace{-0.12cm}\caption{Two possibilities about the slopes of the
nematic transition lines.}\label{Fig_nematic_line}
\end{figure}

To examine how the relation between $g$ and $w$ varies with
decreasing $T$, we have solved the RG equations and obtained its
dependence on the running length scale $l$. To be specific, we have
chosen the following bare values of model parameters: $a^0_s =
-0.001$, $u^0 = 0.05$, $u^0_s = 0.01$, $g^0 = 0.01$, $\lambda^0 =
0.01$, $\lambda^0_{\Delta A} = 1.0 \times 10^{-8}$. We consider
several representative values of $w^0$: $w^0 = 0.1u^0$, $0.3u^0$,
$0.5u^0$, and $1.2u^0$. The $l$-dependence of these parameters can
be easily converted to a $T$-dependence by utilizing the
transformation \cite{Huh2008PRB, She2015PRB, Wang2015PRB} $T = T_0
e^{-l}$, where $T_0$ is some reference temperature smaller than
$T_c$. The numerical results are presented in
Fig.~\ref{Fig_Tem_dependece}(a). We now determine whether the
nematic state becomes a non-nematic state as $T$ is lowered down to
zero on the basis of these results.

\begin{figure}
\centering
\includegraphics[width=3.7in]{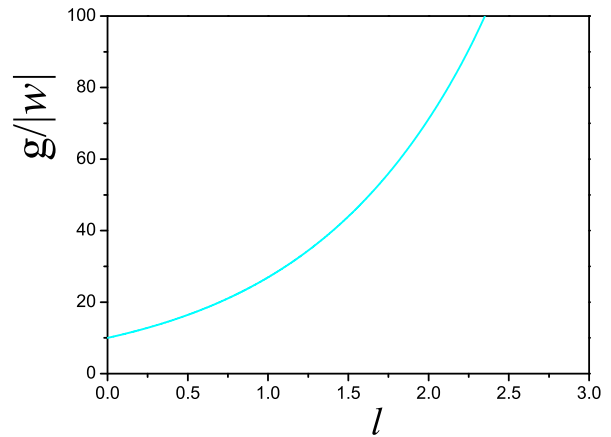}
\vspace{-0.82cm} \caption{(Color online) The evolutions of $g/|w|$
at some representative initial values of parameters: $a^0_s =
-0.001$, $u^0_s = 0.05$, $u^0 = 0.01$, $g^0 = 0.01$, $\lambda^0 =
0.01$, and $\lambda_{\Delta A}^0 = 1.0\times10^{-8}$. The basic
conclusion is independent of these initial values.}\label{Fig_g_w}
\end{figure}

There are in principle two possibilities about the slopes of nematic
transition line $T_{m4}$, as schematically shown in
Fig.~\ref{Fig_nematic_line}. We consider an arbitrary point $A$
lying slightly below the transition line $T_{m4}$. At point $A$, the
system is in the nematic state with $g > \mathrm{max}(0,-w)$. We
then lower $T$ along the route $A \rightarrow B$. If the inequality
$g > \mathrm{max}(0,-w)$ is always satisfied as $T \rightarrow 0$
along $A \rightarrow B$, the system is always in the nematic state
and the slope of the transition line $T_{m4}$ is negative. This
corresponds to the case represented by
Fig.~\ref{Fig_nematic_line}(a). In contrast, if the condition $g
> \mathrm{max}(0,-w)$ is violated as $T$ is reduced to certain
value, the second possibility shown in
Fig.~\ref{Fig_nematic_line}(b) occurs. In this case, the nematic
state becomes non-nematic once again and the slope of transition
line becomes positive at lower temperatures, exhibiting back-bending
behavioe. The numerical results of Eq.
(\ref{Eq_Renormalized_RG_QCP}) informs that the ratio $g/|w| \gg 1$
for various values of $l$, wherein $w$ may be both positive and
negative (it is negative for the curve shown in Fig.~\ref{Fig_g_w}).
From the asymptotic behaviors of $g(l)$ presented in
Fig.~\ref{Fig_Tem_dependece}(a) and $g(l)/|w(l)|$ presented in
Fig.~\ref{Fig_g_w}, we infer that the inequality $g > max(0,-w)$
remains true as $T \rightarrow 0$ if it is satisfied at the starting
point $A$. This clearly indicates that the nematic transition line
$T_{m4}$ has a negative slope inside the SC dome and never bends
backwards, which is well consistent with the observed phase diagram
\cite{Hardy2015NComm}.

\subsection{Suppression of superconductivity due to ordering competition}

We now verify whether superconductivity is suppressed by ordering
competition in the vicinity of the magnetic QCP. To this end, we
will compute the superfluid density after taking into account the
ordering competition among nematic ($C_2$ SDW), $C_4$ SDW and SC
orders. The $T$-dependence of superfluid density $\rho_s(T)$ is
computed based on the $l$-dependence of parameter $\alpha_A \equiv
\alpha_A(l)$ \cite{Huh2008PRB, She2015PRB}, which is extracted from
the coupled flow equation (\ref{Eq_Renormalized_RG_QCP}) and hence
captures the ordering competition. The superfluid density of
superconductor has the generic form $\rho_s(T) = \rho^A_s(T) -
\rho_n(T)$, where $\rho^A_s(T)$ can be evaluated by virtue of the
formula $\rho^A_s(T) \propto \alpha_A(T)$ with $\alpha_A(T)$ being
the mass for vector potential $\mathbf{A}$ generated via
Anderson-Higgs mechanism \cite{Halperin1974PRL} and $\rho_n(T)$ is
the density of thermally excited normal (non-SC) fermionic
quasiparticles.

In this work, we consider only the competition between distinct
order parameters and neglect the contribution of the normal
component, i.e. $\rho_n(T)\ll\rho^A_s(T)$, which is possible for the
$T\ll T_c$, focusing on how superfluid density is modified by
ordering competition. To determine the impact of ordering
competition, we suppose a specific temperature $T_0$ as a reference,
and then examine how superfluid density $\rho_s$ varies as a
function of the ratio $T/T_0$. We assume that $T_0$ is well below
$T_c$ so that the normal fermionic quasiparticles can nearly be
neglected and hence $\rho_s(T)\sim \rho^A_s(T)$. From the results
displayed in Fig.~\ref{Fig_Tem_dependece}(b), we can see that
$\rho^A_s(T)$ is strongly dependent of $T$ in the presence of
ordering competition and decreases rapidly as $T/T_0$ grows. It is
thus clear that the superfluid density is strongly suppressed by
ordering competition and approximately goes to zero in the vicinity
of the point $C$, where $T\approx 0.8 T_0$.

We then consider the impact of ordering competition on $T_c$. The
value of $T_c$ can be determined by solving the equation
$\rho_s(T_c) = \rho^A_s(T) - \rho_n(T_c) = 0$. Although the
contribution $\rho_n(T)$ is not known, we can still infer that $T_c$
is suppressed by ordering competition because $\rho_s$ is
significantly reduced. As shown in Fig.~\ref{Fig_Tem_dependece}(b),
$\rho_s(T)$ vanishes at certain point with $T < T_0$
($T\approx0.8T_0$). This conclusion is well consistent with recent
experiment \cite{Hardy2015NComm}, in which a considerable drop of
$T_c$ is observed near the putative magnetic QCP. In an improved
theoretic treatment, one would compute $\rho_n(T)$ by incorporating
the contribution of fermionic quasiparticles. Notice that these
quasiparticles are not free, but couple strongly to the SDW order
parameter at the magnetic QCP \cite{Liu2012PRB,
Wang_Chubukov2013PRL, Levchenko_Chubukov2013PRL, Chowdhury2013PRL,
Nomoto2013PRL, Kang2015PRB, Chubukov2016PRB}. Usually, this coupling
tends to excite more fermionic quasiparticles out of the SC
condensate, which further suppresses the superfluid density and
reduces $T_c$ \cite{Liu2012PRB, Wang_Chubukov2013PRL,
Levchenko_Chubukov2013PRL, Chowdhury2013PRL, Nomoto2013PRL,
Kang2015PRB, Chubukov2016PRB}.

\begin{figure}
\centering
\includegraphics[width=3.7in]{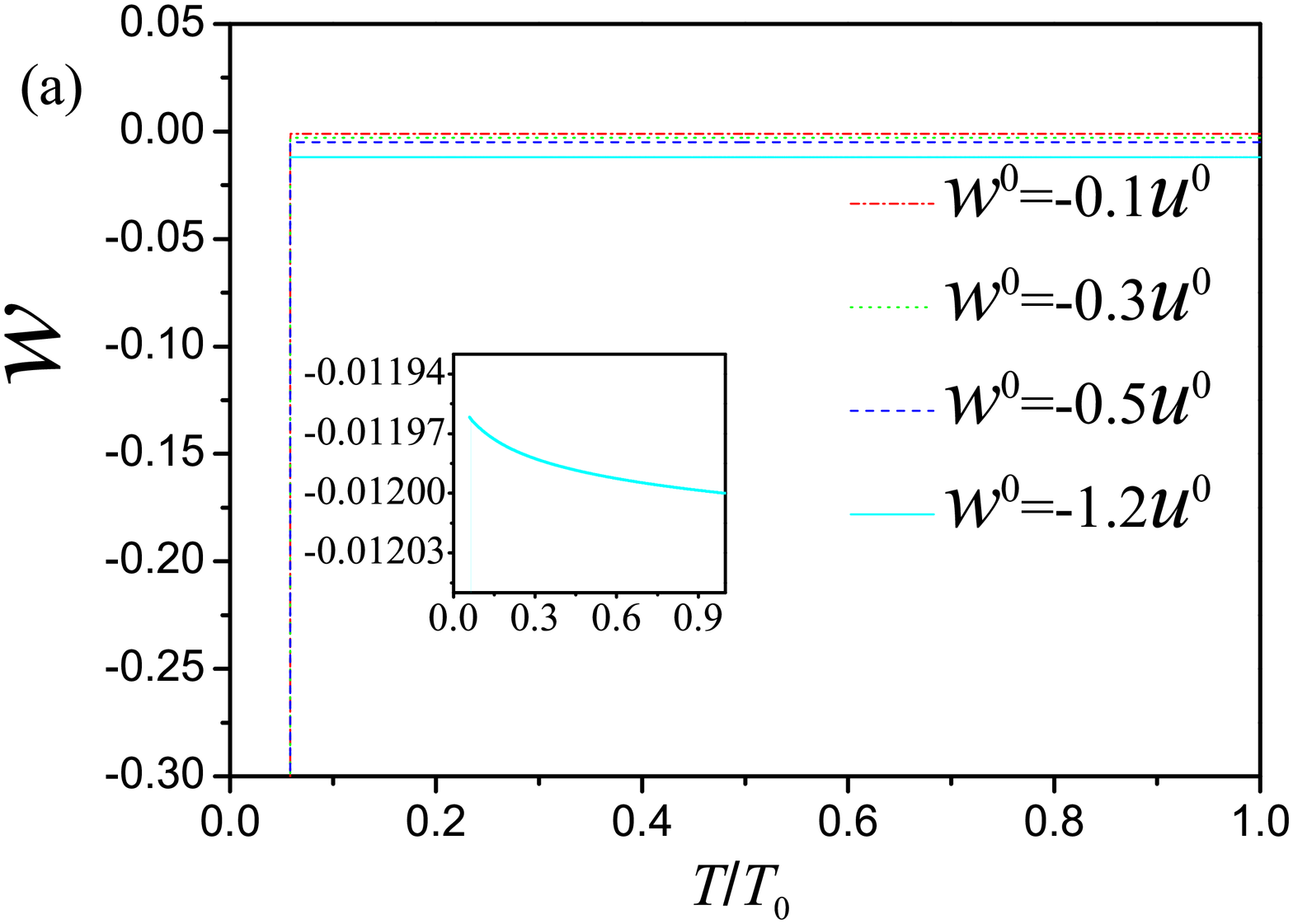}\vspace{-0.6cm}
\includegraphics[width=3.7in]{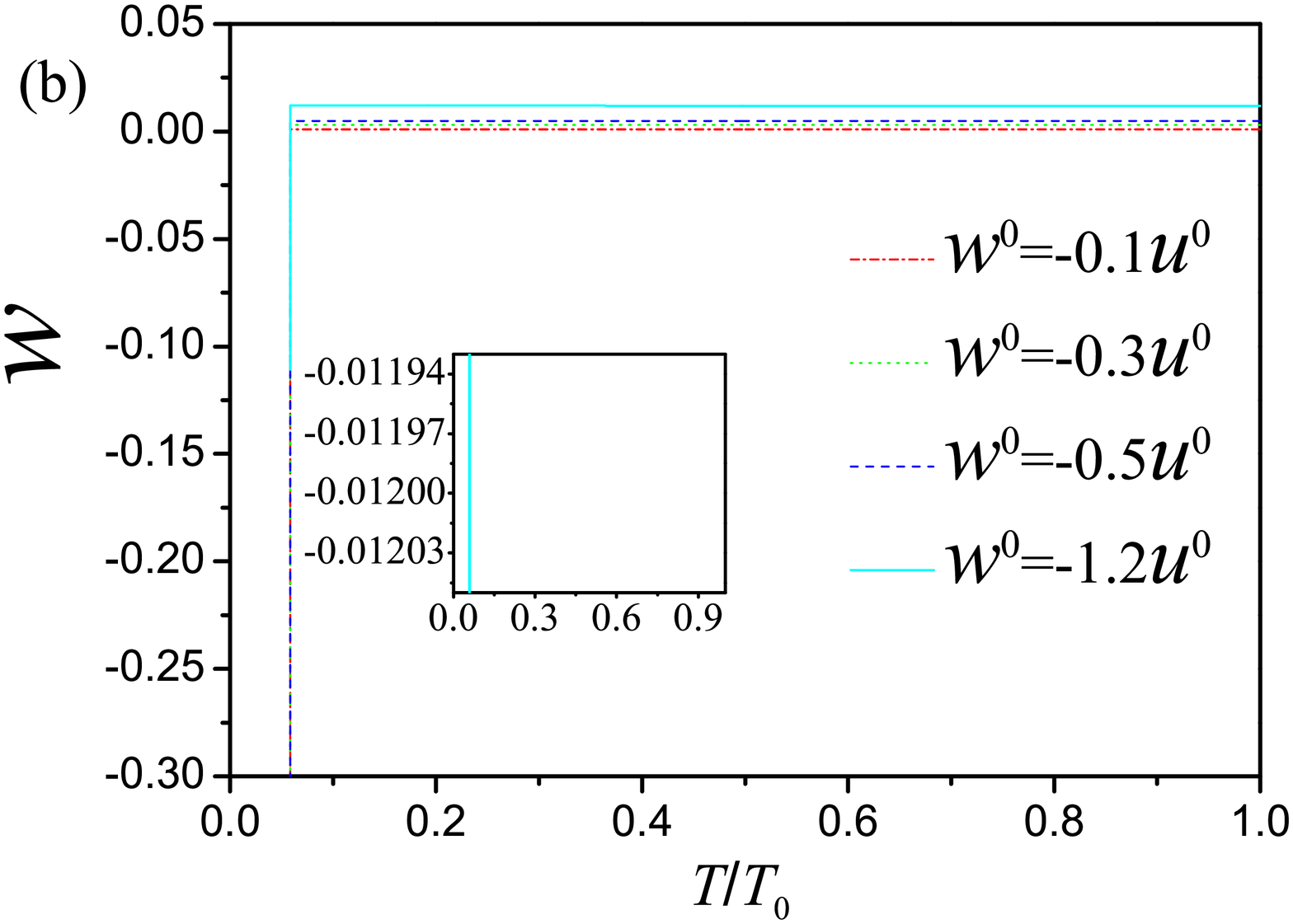}
\vspace{-0.9cm}\caption{(Color online) $T$-dependence of parameter
$w$ obtained at given bare values of dimensionless parameters:
$a^0_s=-0.001$, $u^0_s=0.05$, $u^0=0.01$, $g^0=-0.01$,
$\lambda^0=0.01$, and $\lambda_{\Delta A}^0=1.0\times10^{-8}$. The
main conclusion is independent of these bare values. The parameter
$w$ is not very sensitive to the change of $T$, so we provide two
insets to show more clearly how $w$ varies as $T$ is decreasing. (a)
Starting point is CSDW-type $C_4$ state with $w^0<0$; (b) Starting
point is SVC-type $C_4$ state with $w^0>0$.}
\label{Fig_I_w0_different}
\end{figure}

\subsection{$C_4$ symmetric magnetic order}

We finally turn to analyze the property of $C_4$ symmetric magnetic
state. To uncover the effects caused by ordering competition, we
consider the evolution of the system along the route $D \rightarrow
E \rightarrow F$, shown in Fig.~\ref{Fig_phase_diagram}, and examine
how $g$ and $w$ vary along this route.

Eq.~(\ref{Eq_L}) clearly shows that $w$ is associated with the
quadratic term of $C_4$ SDW order parameter \cite{Fernandes2016PRB,
Schmalian2016PRB}. In analogy to the nematic transition, the sign of
$w$ determines which sort of $C_4$ SDW order, either SVC or CSDW, is
realized \cite{Fernandes2014NPhys, Fernandes2012PRB, Wang2015PRB,
Fernandes2016PRB, Schmalian2016PRB}. In particular, a $C_4$ SVC
order is generated for $g<0$ and $w > 0$, whereas $g < -w$ and $w <
0$ implies the occurrence of a $C_4$ CSDW order
\cite{Fernandes2012PRB, Fernandes2014NPhys, Wang2015PRB,
Fernandes2016PRB, Schmalian2016PRB}. By paralleling the analysis
made for nematic critical line, we convert the $l$-dependence of $w$
using the transformation $T = T_0 e^{-l}$. If one assumes that $w$
is negative at the starting point $D$, which amounts to supposing
the system is in the CSDW-type $C_4$ magnetic state, it remains
negative as $T$ decreases down to the $F$ point, as can be clearly
seen from Fig.~\ref{Fig_I_w0_different}(a). This result implies that
CSDW state is stable in the low-$T$ region. On the other hand, if
one starts from a positive $w$, corresponding to a SVC-type $C_4$
magnetic state, we show in Fig.~\ref{Fig_I_w0_different}(b) that $w$
eventually becomes negative at some critical $T$, which can be
identified as the $E$ point. It follows that the SVC state is
unstable in the low-$T$ region, and that the CSDW state is more
favorable.

We conclude from the above analysis that, although in principle
either SVC or CSDW type $C_4$ state could be realized in the
high-$T$ region, the CSDW-type $C_4$ state is the only stable one in
the low-$T$ region. In a recent work, Christensen \emph{et al.}
\cite{Christensen2015PRB} have suggested the spin-orbit coupling
gives rises to the CSDW-type $C_4$ SDW. Additionally, Hoyer \emph{et
al.} \cite{Schmalian2016PRB} have studied the disorder effects and
demonstrated that impurity scattering favors CSDW over SVC. Here, we
provide a different approach to determine the nature of $C_4$
symmetric magnetic order. Moreover, the sudden drop of $w$ at
certain critical energy scale usually indicates the happening of a
first-order transition, which is qualitatively consistent with
recent experiments \cite{Hardy2015NComm}.

\section{Comparison to $\mathrm{Ba(Fe_{1-x}Co_x)_2As_2}$}\label{Sec_discussions_2}

We now compare the present RG results with a previous work
\cite{Wang2015PRB}, which investigated the impact of ordering
competition on the global phase diagram of
$\mathrm{Ba(Fe_{1-x}Co_x)_2As_2}$. Both
$\mathrm{Ba(Fe_{1-x}Co_x)_2As_2}$ and $\mathrm{Ba_{1-x}K_xFe_2As_2}$
belong to the 122 family of FeSCs, and display a complicated phase
diagram. A common feature is that, over a large part of their phase
diagrams, superconductivity coexists and competes with a SDW type
magnetic order and a nematic order. The ordering competition and its
effects on the phase diagram can be described by deriving an
effective low-energy field theory which is supposed to contain
several distinct order parameters \cite{Fernandes2013PRL,
Fernandes2016PRB, Schmalian2016PRB, Fernandes2017RPP}. Such an
effective theory is expected to be as general as possible, and
applicable in $\mathrm{Ba(Fe_{1-x}Co_x)_2As_2}$,
$\mathrm{Ba_{1-x}K_xFe_2As_2}$, and other similar 122 FeSCs.

However, there are some important differences between the compounds
$\mathrm{Ba(Fe_{1-x}Co_x)_2As_2}$ and
$\mathrm{Ba_{1-x}K_xFe_2As_2}$. In
$\mathrm{Ba(Fe_{1-x}Co_x)_2As_2}$, there is only a $C_2$ symmetric
stripe-type SDW order. In contrast, there are both $C_2$ and $C_4$
symmetric magnetic states in $\mathrm{Ba_{1-x}K_xFe_2As_2}$ and a
number of other hole-doped 122 FeSCs \cite{Goldman2010PRB,
Osborn2014NatureComm, Wang2016PRB, Osborn2016NaturePhys,
Hassinger2012PRB, Hardy2015NComm, Allred2015PRB, Hassinger2016PRB}.
Moreover, the nematic transition line exhibits completely different
doping dependence in the SC dome of these two FeSCs: it has a
positive slope inside the SC dome of
$\mathrm{Ba(Fe_{1-x}Co_x)_2As_2}$ \cite{Nandi2010PRL}, but a
negative slope inside the SC dome of $\mathrm{Ba_{1-x}K_xFe_2As_2}$
\cite{Hardy2015NComm}.

To capture both the similarity and difference, the model of
$\mathrm{Ba_{1-x}K_xFe_2As_2}$ should be formally analogous but not
identical to that of $\mathrm{Ba(Fe_{1-x}Co_x)_2As_2}$
\cite{Wang2015PRB}. It was suggested in Refs.\cite{Fernandes2016PRB,
Schmalian2016PRB} that the $C_4$ symmetric magnetic order that
emerges in $\mathrm{Ba_{1-x}K_xFe_2As_2}$ can be described by
introducing a new term $-\left(\mathbf{M}_X \times
\mathbf{M}_Y\right)^2$. As shown previously in
Ref.\cite{Wang2015PRB}, in the absence of this term, ordering
competition gives rise to the suppression of superconductivity and
in particular the positive slope of nematic transition line in the
SC dome, which are in good agreement with experiments performed in
$\mathrm{Ba(Fe_{1-x}Co_x)_2As_2}$ \cite{Nandi2010PRL}. In the
current work, we have demonstrated through RG calculations that,
adding the above new term leads to the suppression of
superconductivity near the magnetic QCP and also the negative slope
of nematic transition line in the SC dome of
$\mathrm{Ba_{1-x}K_xFe_2As_2}$, which is qualitatively consistent
with recent experiments \cite{Hassinger2012PRB, Hardy2015NComm,
Allred2015PRB, Hassinger2016PRB}. Furthermore, our RG analysis
revealed that the CSDW-type $C_4$ magnetic state is more favorable
than the SVC-type $C_4$ magnetic state, and hence can be used to
determine the nature of $C_4$ magnetic state observed in
$\mathrm{Ba_{1-x}K_xFe_2As_2}$ \cite{Hardy2015NComm}. It is
therefore clear that the effective model of
$\mathrm{Ba(Fe_{1-x}Co_x)_2As_2}$ can be properly modified to
describe $\mathrm{Ba_{1-x}K_xFe_2As_2}$, and that the same
perturbative RG scheme used in Ref.\cite{Wang2015PRB} and here can
be applied to account for both the similarity and difference between
$\mathrm{Ba(Fe_{1-x}Co_x)_2As_2}$ and
$\mathrm{Ba_{1-x}K_xFe_2As_2}$.

\section{Summary and discussion}\label{Sec_summary}

In summary, we have studied the impact of the competition between
superconductivity and $C_2$ and $C_4$ symmetric magnetic orders in a
hole-doped FeSC $\mathrm{Ba_{1-x}K_xFe_2As_2}$. After performing a
detailed RG analysis within an effective field theory, we have
reproduced a number of interesting features of the global phase
diagram. In particular, our RG analysis have showed that the order
parameter fluctuation and ordering competition lead to moderate
suppression of superconductivity near the magnetic QCP, maintain the
negative slope of nematic critical line in the SC dome, and also
sort out the CSDW-type $C_4$ magnetic order as the more stable
state than a SVC-type $C_4$ magnetic order in the low-$T$ regime.
All these theoretic results are well consistent with the recent
experiments of Ref.~\cite{Hardy2015NComm}, and schematically
summarized in Fig.~\ref{Fig_phase_diagram}.

Our RG calculations are confined to the small region surrounding the
magnetic QCP in the SC dome. To gain a better knowledge of the
entire phase diagram, it is necessary to consider the non-SC phase
above $T_c$. A salient feature observed in
Ref.~\cite{Hardy2015NComm} is the back-bending behavior of a
critical line between nematic ($C_2$ SDW) to pure $C_4$ SDW orders,
namely $T_{M4}$, which turns out to be induced by the emergence of
$C_4$ symmetric magnetic order. The transition line $T_{M4}$ exists
well above $T_c$, thus there is no SC order and the competition
between SC and magnetic order parameters is unlikely to be
important. It turns out that the underlying mechanism for the
back-bending behavior of $T_{M4}$ is entirely different from that is
used to account for the slope of $T_{m4}$ in the SC dome. We believe
that an essential role is played by elementary fermionic degrees of
freedom, which are strongly suppressed below $T_c$ by the SC gap but
should be present above $T_c$. The inter-fermion interaction is
expected to be responsible for the transition between $C_4$ and
$C_2$ symmetric magnetic states. This problem is made more
complicated by the uncertainty of the nature of $C_4$ symmetric
magnetic order. In the SC dome below $T_c$, ordering competition
lifts the degeneracy between CSDW and SVC states at low energies,
and chooses CSDW as the true ground state. However, order
competition is much less important above $T_c$. It remains unclear
whether the CSDW or SVC state is realized in the region between
$T_{M4}$ and $T_c$. The microscopic mechanism for the back-bending
behavior of $T_{M4}$ could be properly understood only after the
nature of $C_4$ symmetric magnetic order is identified, which is
subject to future research.

In this paper, we have considered only one specific compound
$\mathrm{Ba_{1-x}K_xFe_2As_2}$. Recent experiments of Hardy \emph{et
al.} \cite{Hardy2015NComm} provided a clear and detailed global
phase diagram of $\mathrm{Ba_{1-x}K_xFe_2As_2}$, which gives us a
good opportunity to directly compare our RG results with
experimental results. Apart from $\mathrm{Ba_{1-x}K_xFe_2As_2}$, the
$C_4$ symmetric magnetic order also exists in a number of other
hole-doped FeSCs, including $\mathrm{Ba(Fe_{1-x}Mn_x)_2As_2}$
\cite{Goldman2010PRB}, $\mathrm{Ba_{1-x}Na_xFe_2As_2}$
\cite{Osborn2014NatureComm, Wang2016PRB}, and
$\mathrm{Sr_{1-x}K_xFe_2As_2}$ \cite{Osborn2016NaturePhys}. It
should be possible to generalize our RG approach to study the global
phase diagrams of these three FeSCs. However, there might be
important difference between $\mathrm{Ba_{1-x}K_xFe_2As_2}$ and
these FeSCs. In that case, the effective field-theoretic model given
by Eq.~(\ref{Eq_L}) needs to be properly modified. Once the modified
effective model is specified, it is straightforward to carry out RG
calculations, just as what we have done in this work.

\section{Acknowledgements}

J.W. and G.Z.L acknowledge the financial support from the National
Natural Science Foundation of China under Grants 11504360, 11274286,
and 11574285. J.W. is also partly supported by the China
Postdoctoral Science Foundation under Grants 2015T80655 and
2014M560510, the Fundamental Research Funds for the Central
Universities (P. R. China) under Grant WK2030040074, and the Program
of Study Abroad for Young Scholar sponsored by CSC (China
Scholarship Council). D.V.E. and J.v.d.B would like to acknowledge
the financial support provided by the German Research Foundation
(Deutsche Forschungsgemeinschaft) through priority program SPP 1458.
J.v.d.B is also supported by SFB 1143 of the Deutsche
Forschungsgemeinschaft.

%%%%%%%%%%%%%%%%%%%%%%%%%%%%%%%%%%%%%%%%%%%%%%%%%%%%%%%%%%%%%%%%%%%

\end{document}